\documentclass[prl,amsmat,amssymb,amsfonts,twocolumn,showpacs]{revtex4}
\usepackage{graphicx}
\usepackage{dcolumn}
\usepackage{bm}

\begin{document}

\author{F. Mu\~noz-Rojas}
\author{J. Fern\'andez-Rossier}
\email{jfrossier@ua.es}
\author{J. J. Palacios}
\affiliation{Departamento de F\'\i sica Aplicada, Universidad
de Alicante, San Vicente del Raspeig, E-03690 Alicante, Spain.}
%%%%%%%%%%%%%%%%%%%%%%%%%%%%%%%%%%%%%%%%%%%%%%%%%%%%%%%%%%%%%%%%%%%%%
%% The document title should be given as usual
%% A short title can be given as a *suggestion* for running headers.
%%%%%%%%%%%%%%%%%%%%%%%%%%%%%%%%%%%%%%%%%%%%%%%%%%%%%%%%%%%%%%%%%%%%%
\title{Giant magnetoresistance in ultra-small Graphene based  devices}

\date{\today}

\begin{abstract}

By computing spin-polarized electronic transport across a finite zigzag graphene ribbon bridging two metallic graphene 
electrodes, we demonstrate, as a proof of principle, that devices featuring 100\% magnetoresistance can be built entirely out of carbon.  In the ground state a short zig-zag ribbon  is an antiferromagnetic insulator which, when connecting two metallic electrodes, acts as a tunnel barrier that suppresses 
the conductance.   Application of a magnetic field turns the ribbon ferromagnetic  and conducting, increasing 
dramatically the current between electrodes. We predict large magnetoresistance in this system  at liquid nitrogen temperature and 10 Tesla or at liquid helium temperature and 300 Gauss. 
 
\end{abstract}

%\pacs{PACS numbers: }

\maketitle

The term spintronics has been coined to refer to the interplay between spin
polarization and electrical conductance and  is one of the major themes today in
condensed matter and applied physics. A central  concept in spintronics is that
of giant magnetoresistance (GMR), discovered originally  in layered structures 
alternating magnetic and non-magnetic  transition metals\cite{GMR}. The
resistance of the whole structure  undergoes a large increase when  the relative
orientation of the magnetization in adjacent layers goes from parallel to
antiparallel,  provided that the non-magnetic layers  are thinner than  the spin
relaxation length. This phenomenon represents the paradigm under which
commercial devices such as magnetic reading heads operate nowadays.

Here we propose a new type of magnetoresistive device which  makes use of the
remarkable electronic properties  of zig-zag graphene
ribbons\cite{Nakada96,Fujita96}. Its operational principle is similar to that
found in conventional GMR layers, with the difference that is entirely based on
carbon.   Our proposal is motivated by the  spectacular progress in the 
fabrication of high mobility graphene based field effect
transistors\cite{Geim05,Kim05,Science06}  and by the recent developments in the
fabrication of graphene ribbons\cite{Natmat,Avouris,kim07,Pablo07} with top-down
techniques,  as well as in the sinthesys of graphene ribbons\cite{Dai08}. 

The electronic structure of infinite graphene ribbons has been studied
thoroughly. Idealized graphene ribbons, with boundaries parallel to the
crystallographic directions and boundary carbon atoms passivated with a single
hydrogen atom, fall into two categories: armchair and zig-zag. Within the
simplest one orbital tight-binding description\cite{Nakada96},  armchair ribbons
can be either semiconducting or metallic, depending on their width, whereas in
the case of zigzag ribbons  two  flat bands, associated with edge states, lie
at  the Fermi energy. These edge flat bands favor the appearance of
magnetization on the edges when  electron-electron repulsion is included in the
calculation, either with a Hubbard model\cite{Fujita96,JFR08,Tomanek08}
 or with density functional
theory (DFT)\cite{Son06,Cohen06,pisani}. In the ground state the respective  magnetization
direction of the edges is antiparallel, and a gap opens in the band
structure\cite{Son06,Cohen06,pisani,JFR08}. This is the ground state. Slightly above
in energy, the parallel magnetic configuration is conducting. Application of
either a magnetic field or a transverse electric field\cite{Cohen06} can make
the ferromagnetic configuration more stable. Here we explore the former.

\begin{figure}
[b]
\includegraphics[width=3.in]{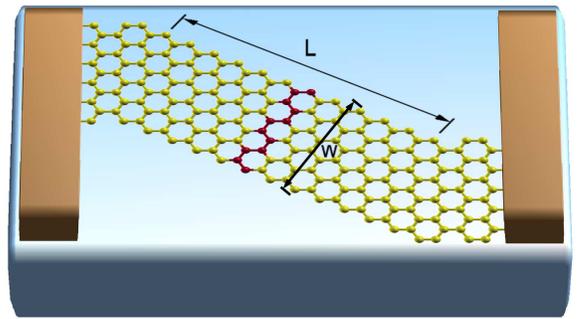}
\caption{ \label{Figure1} Atomic structure of the  zigzag ribbon  with length $N_x=12$ and width $N_y=6$ attached to semi-infinite electrodes. The unit cell of a zigzag ribbon is highlighted.}
%In the upper panel we highlight the unit cell of a zigzag ribbon.  The vector field denotes the magnetization, calculated self-consistently. The magnetic moments are localized in the edges of the zigzag ribbon.  Both the ferromagnetic and antiferromagnetic solutions are shown.      }
\end{figure}

Both for conceptual and practical reasons our proposal is based on finite length $L$
graphene ribbons. For $L=\infty$,  the long range
order predicted by mean field calculations is not robust, due  to the
proliferation of spin wave excitations of energy $L^{-n}$, with $n=1$ or $n=2$
for both antiferromagnetic and ferromagnetic alignments, respectively\cite{Yazyev07}. In
short ribbons, in contrast, there is a gap for spin waves. Quantum fluctuations
between the manifold of equivalent mean field ground states would not alter the
conducting properties of the system. Additionally, recent DFT
calculations\cite{Mauri08} show that monohydrogenated zigzag ribbons are not the
most stable edge
configuration unless the hydrogen density is very small. Whereas this poses a
severe problem for the chemical stability of infinite monohydrogenated zigzag
ribbons, short magnetic ribbons, as well as other magnetic polycyclic aromatic
hydrocarbons\cite{Wu07},  might be more stable.
%when attached to structures wich are
%more efficient adsorbing hydrogen.   

Here we show that the spin driven metal
insulator transition, predicted so far for infinite  ribbons,
%and short stripes,
  is still present in short ribbons attached to conducting electrodes. Using 
a well-established methodology\cite{Munoz06} extended to account for electron-electron
interactions in a Hubbard model, we  study both  the
magnetic  and transport properties of a system in which two conducting graphene
electrodes are coupled through a finite length graphene ribbon.

{\em Statement of the problem and methodology.-} We study the system depicted in
Fig. 1: A short zig-zag graphene ribbon attached to two conducting graphene electrodes.
In the case pictured in the figure we have chosen the electrodes to be semi-infinite metallic
armchair graphene ribbons of width $W$. Both the device and the electrodes are
described with a Hubbard model for the $\pi_z$ orbitals with first-neighbor
hopping $t$ and on-site repulsion $U$, solved at a collinear 
mean field level\cite{Fujita96,JFR07,JFR08,JJP08,Tomanek08,Yazyev08}
This approach is known to capture
the main features of the ab-initio calculations, both for finite\cite{JFR07} and
infinite\cite{JFR08} graphene systems. Thus, the mean-field Hamiltonian reads
\begin{equation}
{\cal H}= \sum_{I,I',\sigma} t c^{\dagger}_{I \sigma} c_{I' \sigma} +
U\sum_{I} \left( n_{I,\uparrow}\langle n_{I,\downarrow} \rangle + n_{I,\downarrow} \langle n_{I,\uparrow} \rangle \right)
%{\cal H}= \sum_{\vec{r},\vec{r}',\sigma} t_{\vec{r},\vec{r}'}c^{\dagger}_{\vec{r}\sigma}c_{\vec{r}'\sigma} + U\sum_{\vec{r}}n_{\vec{r},\uparrow}\langle n_{\vec{r},\downarrow} \rangle + n_{\vec{r},\downarrow}\langle n_{\vec{r},\uparrow} \rangle
\label{HMF0}
\end{equation}
where $c^{\dagger}_{I\sigma}$ creates an electron at the $\pi_z$
orbital of atom $I$ with spin $\sigma$,
$n_{I,\sigma}=c_{I \sigma}^{\dagger}c_{I \sigma}$ is the
occupation operator.  Since the mean fields 
$\langle n_{I,\sigma} \rangle$ depend on the eigenstates of 
the mean field Hamiltonian, this defines a self-consistent problem 
which is solved by numerical interation. The converged solutions define a
 one body Hamiltonian for the electrons in the structure.
%%%%%%%%%%%%%%%%%%%%%Description of the methodology

In this work we have  to solve the self-consistent problem for  an infinite
system without translational invariance. This is done using the partition method\cite{Munoz06}. 
We split the system in three sectors, the left and right electrodes
and the central region. The Hamiltonian of Eq. (\ref{HMF0}) reads:
\begin{equation}
{\cal H}= {\cal H}_L +{\cal H}_R + {\cal H}_C  + {\cal V}_{LC} +{\cal V}_{LR} 
\end{equation}
where ${\cal H}_{(L,R,C)}$ are the mean field Hubbard Hamiltonians of the left
and right electrodes and the central region and ${\cal V}_{(R,L)C}$ describes
the hopping between the central region and the electrodes. Sufficiently away
from the central region, the mean field Hamiltonian of the electrodes is
identical to that  of an infinite ribbon. In the case of the armchair
electrodes, the effect of the Hubbard interactions in the charge neutrality
point is a rigid shift of the bands without the appearance of  magnetic moment
and keeping their metallic character. 
%An important aspect of this method is the choice of the boundary between central region and electrodes.
The first step is the determination of the surface Green function of the semiinfinite electrodes, $g_{(L,R)}$ \cite{Munoz06}. This requires the solution of a self-consistent Dyson equation. Once this is done, the Green function of the central region reads:
\begin{equation}
G_C(E) \equiv \left[ {EI-{\cal H}_C - \Sigma_L(E) -\Sigma_R(E)} \right]^{-1}
\label{Greendevice}
\end{equation} 
with $\Sigma_{(L,R)}(E)= {\cal V}_{(L,R)} g_{(L,R)}(E) {\cal V}^{\dagger}_{(L,R)}$
%\begin{equation}
%\Sigma_{(L,R)}(E)= {\cal V}_{(L,R)} g_{(L,R)}(E) {\cal V}^{\dagger}_{(L,R)} 
%\label{Self}
%\end{equation} 

This expression is a functional of the  expectation values $\langle
n_{I,\sigma} \rangle$, through the central Hamiltonian ${\cal H}_C$.  The
Green function yields the  density of states projected over the orbital $\pi_z$
with spin $\sigma$ sitting in the atom $I$ in the device, is given by:
\begin{equation}
\rho(E,I,\sigma) \equiv -\frac{1}{\pi}
 Im {\rm Tr}\left(\langle I\sigma |G_C(E)|I\sigma\rangle \right)
\label{localdos}
\end{equation} 
Here $\langle I\sigma |G_C(E)|I\sigma\rangle$ is a matrix of
size $2N$, with $N$ being the number of atoms of the central region. In turn,
the expectation values of the spin density  are given by 
% $\langle n_{I,\sigma} \rangle = \int_{-\infty}^{E_F} \rho(E,I,\sigma)dE$

\begin{equation}
\langle n_{I,\sigma} \rangle = \int_{-\infty}^{E_F} \rho(E,I,\sigma)dE
\label{densigreen}
\end{equation} 

The magnetic moment in a given atom I is  defined as $m_I= \frac{\langle n_{I,\uparrow} \rangle-\langle n_{I,\downarrow} \rangle}{2}$
%\begin{equation}
%m_I= \frac{\langle n_{I,\uparrow} \rangle-\langle n_{I,\downarrow} \rangle}{2}
%\end{equation}

%\begin{equation}
%\langle n_{\vec{r},\sigma} \rangle= -\frac{1}{\pi}\int_{-\infty}^{E_F}
% Im {\rm Tr}\left(\langle \vec{r}\sigma |\hat{G}_C(E)|\vec{r}\sigma\rangle \right)dE
%\label{densigreen}
%\end{equation} 

Equations (\ref{Greendevice}) and (\ref{densigreen}) define a self-consistent
problem which is solved by numerical iteration. The solution provides a mean
field description of the central region attached to the electrodes.  
%The boundaries between the central region and the electrodes are chosen so that the electronic structure in the electrodes is not affected by the electronic structure of the device. From this point of view, it is convenient to include in the central region a portion of the electrode as large as possible with the resulting computational overhead. The optimal choice is the one for which no change in the final self-consistent result is obtained when the central regionis increased in size. 
\begin{figure}
[hbt]
\includegraphics[width=3.in]{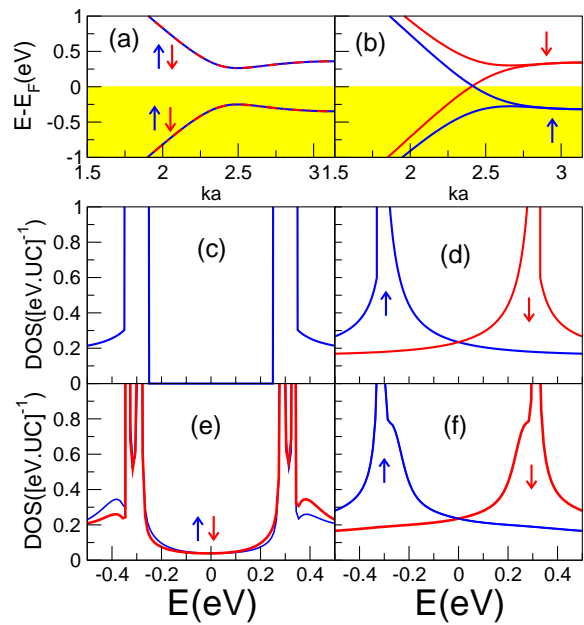}
\caption{ \label{Figure2} (a,b) Bands for infinite ribbon with $N_y=12$
with AF (a) and FM (b) ground states. (c,d)  
Corresponding DOS of the AF (c) and FM (d) infinite ribbons. 
(e) DOS of the AF finite ribbon. 
 (f) DOS of the FM finite ribbon. }
\end{figure}
Within this formalism the Landauer conductance is given by $G=\frac{e^2}{h} {\rm
Tr}T(E_F)$ with:
\begin{equation}
T(E)= {\rm Tr}\left( \Gamma_L(E) G_C \Gamma_R(E) G^{\dagger}_C(E)\right)
\end{equation}
with $\Gamma_L(E)= i\left(\Sigma_{(L,R)}(E)-\Sigma^{\dagger}_{(L,R)}(E)\right)$
%\begin{equation}
%\Gamma_L(E)= i\left(\Sigma_{(L,R)}(E)-\Sigma^{\dagger}_{(L,R)}(E)\right)
%\end{equation}

{\em Results.-} 
%We study the electronic structure of finite length zigzag ribbons attached to
%conducting armchair ribbons. 
The size of the zig-zag ribbons is defined by two integer
numbers, $N_x$, the number of unit cells of the ribbon, and $N_y$ the number of
zigzag rows in the unit cell. Thus, the length of the ribbon is $L=N_x a$, with
$a=2.42$ \r{A} and the width of the ribbon is given by $W=\frac{\sqrt{3}}{2}N_y a$.

We verify first that magnetic solutions appear in short zigzag ribbons connected
to conducting electrodes.  As in the case of both isolated finite ribbons\cite{Scuseria}
and infinite ribbons\cite{Cohen06}, we obtain two
kind of solutions, ferromagnetic and antiferromagnetic,
depending on the initial guess in the iterative self-consistent procedure. We
see how only the edge atoms of the zigzag ribbon are magnetic and how the size
of the edge  moments is larger away from the interface. In the case of the AF
solution, the largest magnetic moment is  $m_I=0.13$  for $U=t=2.7$eV almost
identical to the result of infinite ribbons.  In Fig. 2 we   compare the
electronic structure of the finite size connected zigzag ribbons with the case
of infinite ribbons, both for ferromagnetic (FM) and antiferromagnetic (AF)
configurations.  The electronic structure of the  infinite ribbons is calculated
taking advantage of  crystal invariance and making use of the  Bloch theorem\cite{JFR08}. %
Whereas AF infinite ribbons  have a gap $\Delta_0$\cite{Son06,JFR08}
and zero density of states (DOS) at $E_F$ (hereon set to zero),
two bands cross $E_F$ for the FM ribbons, resulting in
a finite density of states at $E_F$.
The same trend is observed in finite ribbons. The DOS of the AF
configurations, summing over the atoms of a unit cell (see Fig. 2),  presents a
pseudogap. The DOS becomes more similar to the one of the infinite case both
for longer ribbons and for unit cells in the middle of the ribbon. The  small
DOS inside the gap arises from the coupling to the electrodes. 
%This is shown in the inset where we plot the density of states at $E=E_F$,
%summed over the two edge atoms, as a we cross the ribbon from one electrode to
%the other. 
%It is apparent how the central region has a smaller LDOS, more
%similar to the infinite ribbon solution.
    In contrast to the AF solution, the
DOS of the FM solutions,  again summing over the atoms of a unit cell
(see Fig. 2), is that of a conducting system.  

\begin{figure}
[hbt]
\includegraphics[width=3.in]{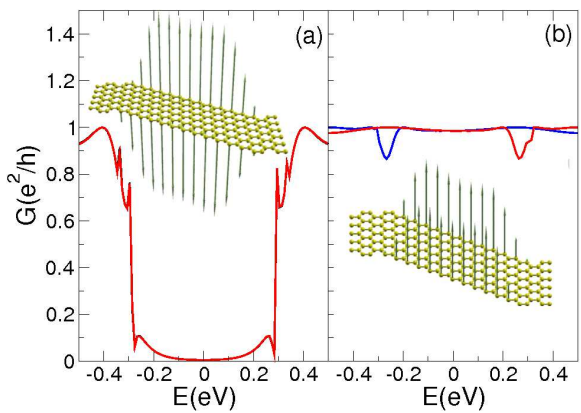}
\caption{ \label{Figure3} (a) Spin resolved transmission for the  AF infinite
ribbon with $N_x=12$ and $N_y=6$. (b) Same for the  FM solution. }
\end{figure}

From the DOS we anticipate a strong dependence of the conductance on the magnetic configuration.
Our calculations confirm this expectancy. In Fig. 3 we show the
conductance curves for the $N_x=12, N_y=6$ ribbon. In the left panel we show how
the transmission of the AF solution for $E=E_F$ is highly reduced, compared to
the FM solution, shown in the right panel. This is the central result of our
work: The conductance is strongly dependent on the
relative orientation of the magnetization of the edges of the zigzag ribbon bridging the electrodes. 
 In the ground state the
ribbon is antiferromagnetic and the conductance is small or vanishes for
sufficiently large $N_x$.  In Fig. 4 we plot the conductance at $E_{F}$, both for AF and
FM configurations,     as a function of the ribbon length $L=aN_x$ for two widths.
Only for very short ribbons the conductance in the AF case is not vanishingly
small, due to tunneling. The conductance of the FM state is always close to
$2e^2/h$, value associated with the two  bands, one per spin channel,  that cross the
Fermi energy. This conductance value is sensitive to the actual form of the graphene electrodes, but
the vanishingly small values for the AF state are not. 

As in standard GMR devices,
the application of a magnetic field can force the ferromagnetic solution to be the ground
state, resulting in a dramatic increase of the conductance.   We define the 
magnetoresistance (MR) of each device as
\begin{equation}
{\rm MR}\equiv \frac{R_{AF}-R_{FM}}{R_{AF}+R_{FM}}\times 100
=\frac{G_{FM}-G_{AF}}{G_{FM}+G_{AF}}\times 100,
\label{MR}
\end{equation} 
where $R=\frac{1}{G}$ is the resistance and $G=\frac{e^2}{\hbar}T(E_F)$ is the
conductance calculated with the Landauer formula. In Fig. 4 we plot the MR as
a function of the ribbon length for ribbons of width $N_y = 6$ and $N_y = 3 $.  We see
that, except for ribbons shorter than 1nm, the MR saturates to 100$\%$. 
The origin of the MR proposed here is different from that discussed by Brey and
Fertig\cite{BreyMR} and that of Kim and Kim\cite{kim08} which require
ferromagnetic electrodes. Our proposal is more similar to the original GMR
experiments with current in plane in exchanged coupled multilayers\cite{GMR}. 

\begin{figure}
[hbt]
\includegraphics[width=3.in]{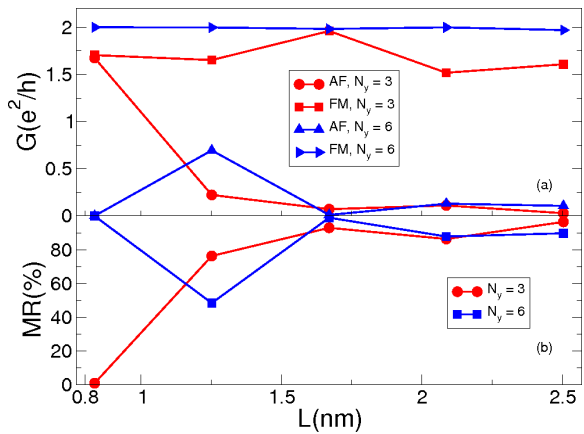}
\caption{ \label{Figure4} (a) Conductance and (b) magnetoresistance (Eq. \ref{MR})  at the Fermi energy as a function of the ribbon length for two ribbon widths, both for FM and AF configurations.}
\end{figure}

We now discuss the operational limits of our proposed device. There are three relevant
magnitudes to be considered. First, the  energy difference between the AF and the
FM states, which depends both on the width of the ribbon and the length:
\begin{equation}
\Delta(N_x,N_y)\simeq N_x\left(\frac{30}{N_y}\right)^{3/2}.
\label{EF-EAF}
\end{equation}
This expression has been obtained from the fitting to DFT calculations by Pisani et al.\cite{pisani}
Second, the temperature-dependent spin correlation length $\xi(T)$, over which magnetic 
order is lost\cite{Yazyev07}. Finally, and neglecting the influence of the magnetic
field on the orbital part of the wavefunctions, the critical switching magnetic field is defined through the equation
\begin{equation}
g\mu_B M_T B^* = \frac{\Delta}{N_x},
\label{Bcrit}
\end{equation} 
where $M_T=\sum_I m_I $ is the total spin density of the FM configuration,
$g=2$ for graphene and $\mu_B=0.058$ meV T$^{-1}$. 

The temperature determines the minimal value $\Delta_{min}$ below which thermal fluctuations will make the AF unstable.
It also determines the maximal value for the length of the ribbon $L_{max}$ above which magnetic order is lost.
By choosing $L=L_{max}$, we guarantee the maximum possible width $W_{max}$ for the ribbon
through Eq. \ref{EF-EAF}. This, in turn, provides the minimum critical switching field $B^*$ through Eq. \ref{Bcrit}.
For instance, at room temperature, $\Delta_{min} \approx 26$ meV and
$\xi(300)=L_{max} \approx 0.8 $nm. This yields $W_{max}\approx 1.6 $nm and $B^*\approx200$ T. 
For liquid nitrogen temperatures (75 K)
$\Delta_{min} \approx 6$ meV, $L_{max} \approx 4 $nm, $W_{max}\approx 12$ nm, and $B^*\approx 10$ T. 
At He liquid temperatures 
(4 K) $\Delta_{min} \approx 0.35$ meV, $L_{max} \approx 80 $nm, $W_{max}
\approx 600$ nm, and $B^*\approx 0.03$ T.
Using the extrapolation formula of Son {\em et al.}\cite{Son06}  for the ribbon transport gaps
a ribbon of $W\approx 600$ nm has a gap of $1.55$ meV, which still larger than $kT$ and standard bias voltages.
In summary, there is a trade-off between the minimum temperature below which the AF ground state becomes unstable
and the critical switching field. This trade-off is controlled by the spin correlation length. At high
temperatures $B^*$ becomes prohibitively large. At liquid He temperatures, the critical field is bounded
within reasonable ranges easily attainable in the lab.

%Our results do not change qualitatively if the electrodes are wider graphene ribbons or even bulk graphene, or other conducting material. In particular,  our conclusions are not affected qualitatively by the fact that "metallic" ribbons do actually have a small gap when long-range interactions are taken into account.
%
%since, for the widths considered,
%this is much smaller than any other relevant energy scale in the problem. 
Also the width of the electrodes does not need to
be equal to the width of the zigzag channel. This would only introduce some additional scattering in the FM state, but never
compromising the high conductance values.

In conclusion, we propose an ultra-small and
chemically simple magnetoresistive device based on a zigzag ribbon joining metallic graphene electrodes.
The conduction properties of this device change
dramatically as the relative orientation of the magnetic edges of the ribbon go from parallel
to antiparallel relative orientations upon application of a magnetic field. 
Even if from the operational point of view there are still challenges regarding room-temperature performance in small
magnetic fields, a proof of principle has been demonstrated.
From a more fundamental point of view, low-temperature experiments showing a drastic change in the resistance on applying 
strong magnetic fields would unambiguously signal the existence of magnetism in zig-zag graphene ribbons.

We acknowledge fruitful discussions with D. Soriano.
This work has been financially supported by MEC-Spain (Grant Nos.
MAT07-67845 and CONSOLIDER CSD2007-0010).

\end{document}